\documentclass[12pt]{iopart}

\usepackage{cite}
\usepackage{iopams}
\usepackage{graphicx}
\usepackage{xcolor}
\usepackage{upgreek}
\begin{document}

\title[The metastable Q $^3\Delta_2$ state of ThO]{The metastable Q $^3\Delta_2$ state of ThO: A new resource for the ACME electron EDM search}

\author{X. Wu$^{1,2}$, Z. Han$^1$, J. Chow$^1$, D. G. Ang$^2$, C. Meisenhelder$^2$, C. D. Panda$^2$\footnote{Present address: Department of Physics, University of California, Berkeley, CA, USA}, E. P. West$^2$\footnote{Present address: Department of Physics and Astronomy, UCLA, Los Angeles, CA, USA.}, G. Gabrielse$^3$, J. M. Doyle$^2$, and D. DeMille$^1$}

\address{$^1$ Department of Physics, Yale University, New Haven, Connecticut 06511, USA}
\address{$^2$ Department of Physics, Harvard University, Cambridge, Massachusetts 02138, USA}
\address{$^3$ Center for Fundamental Physics, Northwestern University, Evanston, Illinois 60208, USA}
\ead{xing.wu@yale.edu}
\vspace{10pt}
\begin{indented}
\item[]\today
\end{indented}

\begin{abstract}
The best upper limit for the electron electric dipole moment was recently set by the ACME collaboration. This experiment measures an electron spin-precession in a cold beam of ThO molecules in their metastable $H~(^3\Delta_1)$ state. Improvement in the statistical and systematic uncertainties is possible with more efficient use of molecules from the source and better magnetometry in the experiment, respectively. Here, we report measurements of several relevant properties of the long-lived $Q~(^3\Delta_2)$ state of ThO, and show that this state is a very useful resource for both these purposes. The $Q$ state lifetime is long enough that its decay during the time of flight in the ACME beam experiment is negligible. The large electric dipole moment measured for the $Q$ state, giving rise to a large linear Stark shift, is ideal for an electrostatic lens that increases the fraction of molecules detected downstream. The measured magnetic moment of the $Q$ state is also large enough to be used as a sensitive co-magnetometer in ACME. Finally, we show that the $Q$ state has a large transition dipole moment to the $C~(^1\Pi_1)$ state, which allows for efficient population transfer between the ground state $X~(^1\Sigma^+)$ and the $Q$ state via $X\textendash C\textendash Q$ Stimulated Raman Adiabatic Passage (STIRAP). We demonstrate $90\,$\% STIRAP transfer efficiency. In the course of these measurements, we also determine the magnetic moment of $C$ state, the $X\rightarrow C$ transition dipole moment, and branching ratios of decays from the $C$ state.
\end{abstract}

%
%
%
%
%

\section{\label{sec:level1}Introduction}

The best upper limit for the electron electric dipole moment (EDM), $d_e$, was recently set by the ACME collaboration: $|d_e|<1.1\times10^{-29}\,$e$\cdot$cm~\cite{ACMECollaboration2018}. This represents an order of magnitude improvement compared to the previous best results~\cite{Baron2014,Baron2017,Cairncross2017}. The ACME limit on $d_e$ significantly constrains the parameter space for potential new time-reversal symmetry ($T$) violating interactions mediated by particles in the mass range of $3\sim30\,$TeV~\cite{Cesarotti2019}. Improved measurements will probe even higher energy scales.


The ACME experiment is based on measurement of an electron spin precession in thorium monoxide (ThO) molecules. These molecules are prepared in electron spin superposition states in the metastable $H~(^3\Delta_1)$ level, where the interaction of an electron EDM is greatly amplified by the effective intra-molecular electric field~\cite{Denis2016,Skripnikov2016} $\mathcal{E}_{\rm eff} \approx 78\,$GV/cm. In order to improve the sensitivity to $d_e$ in ACME, it is necessary to reduce both the statistical uncertainty (now limited by shot noise in the signal from detected molecules~\cite{Panda2019}) and the systematic uncertainty (partially limited by uncontrolled magnetic field gradients~\cite{ACMECollaboration2018}).

The statistical sensitivity can be enhanced by increasing the flux of ThO incident on the downstream detection region. A plausible means to achieve this is through molecular beam focusing. 
Focusing the ground $X~(^1\Sigma^+)$ state using an electrostatic lens has been discussed~\cite{Vutha2010}, but its Stark shift is comparatively weak, so the focusing is not very effective. One could use the $H$ state instead, which has much larger Stark shifts~\cite{Vutha2011}. However, its lifetime is in the millisecond range~\cite{Vutha2010} and this would lead to radiative loss over the beam focusing distance ($\sim\!0.5\,$m). Furthermore, efficient population transfer into $H$ at the lens entrance, corresponding to a much larger phase space area than that in the previous ACME experiment, does not appear feasible with available laser powers.



To improve the systematic uncertainties associated with magnetic field gradients, it would be useful to employ a more accurate magnetic probe. An ideal scenario here is to use a magnetically sensitive state of the ThO molecule as a co-magnetometer, directly at the location where the electron EDM measurement is performed. However, neither the $X$ nor the $H$ state has a large enough magnetic dipole moment for this purpose.

Both these issues could be addressed by using a very long-lived auxiliary state of ThO that possesses both a large Stark shift and a large magnetic moment. To be useful in practice, population transfer into and out of such an auxiliary state should also be efficient. The metastable $Q~(^3\Delta_2)$ state in ThO is expected to satisfy all these requirements simultaneously. In particular:
\begin{figure}
	\centering
		\includegraphics[width=0.68\textwidth]{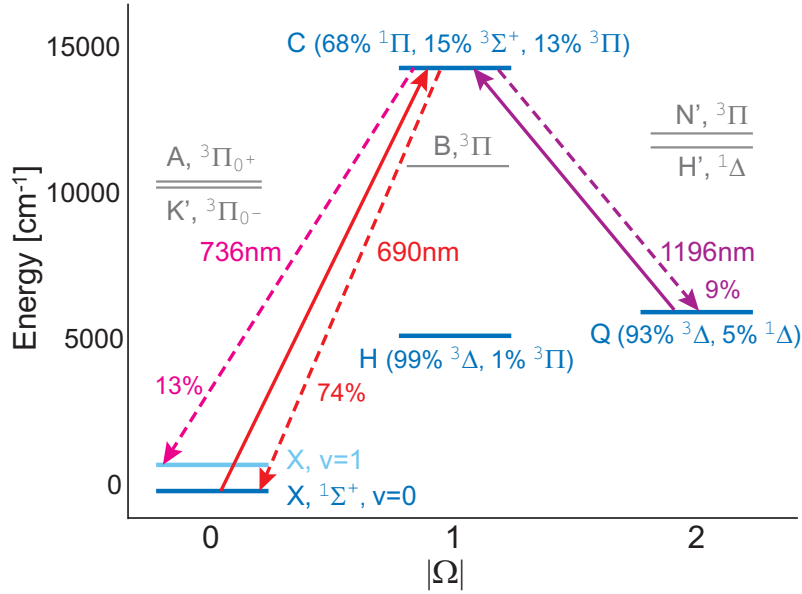}
	\caption{Lowest energy levels of ThO. The horizontal axis shows the $\Omega$ quantum number ($\Omega=\Lambda+\Sigma$, where $\Lambda$ and $\Sigma$ are the projection of total orbital angular momentum $L$ and total electronic spin $S$ on the internuclear axis, respectively). Levels relevant to the current work are indicated in blue. Solid arrows indicate the optical excitations in this work, and dashed arrows indicate the relevant spontaneous decays with their wavelengths and branching ratios labelled nearby. Next to the electronic state designations are the estimated leading contributions from electronic Hund's case (a) terms~\cite{Kaledin2019}. States shown are all $v=0$ vibrational levels unless otherwise noted.}
	\label{fig:levelscheme}
\end{figure}
\begin{itemize}
\item The lifetime of the $Q$ state is expected to be long. Since only two electronic states lie lower in energy (Fig.~\ref{fig:levelscheme}),  it can only decay to the $H$ or $X$ state. The decay to $X$ is doubly forbidden since it requires changing $\Omega$ by $2$ units (i.e. an electric quadrupole transition) and changing the total electronic spin $S$. The transition from $Q$ to $H$ is also nominally forbidden, since $\Sigma = -1$ in the $H\, ^3\Delta_1$ state ($\Sigma$ is the projection of $S$ on the internuclear axis) while $\Sigma = 0$ in the $Q\, ^3\Delta_2$ state, and electric dipole selection rules enforce $\Delta\Sigma = 0$~\cite{Bernath2005}.  However, because of spin-orbit mixing, the Hund's case (a) term symbols for the $Q$ and $H$ state do not exactly describe the state. In particular, the $H$ state includes $\sim1\%$ admixture of $^3\Pi_1$ character, which allows a weak electric dipole matrix element between $Q$ and $H$. The $Q\rightarrow H$ decay rate is further suppressed by its small frequency: the $H$ state is merely $811\,$cm$^{-1}$ below the $Q$ state, and the decay rate scales as the frequency cubed. Based on these facts, the order-of-magnitude of the lifetime is estimated to be~\cite{Bernath2005} 
\begin{equation}
\tau_Q\approx\frac{3\times 10^6}{\tilde{\nu}^3d^2}\left[\textrm{s}\left(\textrm{cm}^{-1}\right)^3\textrm{D}^2\right]\sim0.6\,\textrm{s},
\end{equation}
where the transition frequency $\tilde{\nu}$ is in wavenumber (cm$^{-1}$) and the transition dipole moment $d$ is in Debye (D), with its value estimated by the product between the amplitudes of the Hund's case (a) terms connected by electric dipole transition. (No lifetime measurement on the $Q$ state of ThO prior to this work has been found in the literature.) 
This estimated lifetime would be as good as infinite for the ACME experiment, which requires $\lesssim\!10\,$ms total time of flight, including both the spin precession region and a possible molecular lens.

\item The $Q$ state should be easily polarizable by electric fields. It possesses a pair of nearly degenerate $\Omega$-doublet states, with opposite parities split by less than $\sim\!10\,$kHz~\cite{Edvinsson1984,Edvinsson1985a}. This means the $Q$ state should exhibit a linear Stark shift for any appreciable electric field, and can be deflected strongly in large field gradients. 

\item The $Q$ state is expected to have a substantial magnetic moment of $\approx\!2\mu_B$. It is nominally a $^3\Delta_2$ state, with $\Lambda=2$ and $\Sigma=\Omega-\Lambda=0$. Hence its molecule-frame $g$ factor is $g_Q\approx g_L\Lambda=2$, which is significantly larger than that of the ground $X\,(^1\Sigma^+)$ state ($g_X < 0.001$)~\cite{Eshbach1952} and of the $H\, (^3\Delta_1$) state ($g_H=8.8\times10^{-3}$)~\cite{Kirilov2013}.

\item It should be easy to prepare and read out population in the $Q$ state via the $C~(^1\Pi_1)$ state. $C$ is estimated to possess $13$\% $^3\Pi_1$ character (Fig.~\ref{fig:levelscheme}), to which a dipole transition from $Q$ ($^3\Delta_2$) is allowed.
\end{itemize}

In this paper, we report measurements of the lifetime and the electric and magnetic dipole moment of the $Q$ state in ThO, and demonstrate efficient population transfer between $X$ and $Q$ states via the stimulated Raman adiabatic passage (STIRAP) process. We measure a lower bound for the $Q$ state lifetime, which guarantees negligible spontaneous decay during the entire ACME beam flight path. We show that the $Q$ state has strong linear Stark and Zeeman shifts that allow efficient molecular beam focusing by either an electro- or a magnetostatic lens. The large Zeeman effect also renders the $Q$ state useful for sensitive {\it in situ} co-magnetometry. We measure a strong transition dipole moment for $Q\rightarrow C$ which allows for the demonstration of $90$\% STIRAP efficiency between $X\leftrightarrow Q$ via the intermediate $C$ state, with near-saturation of the population transfer over a broad Doppler distribution and spatial extent of the ThO beam. This paves the way for applying the $Q$ state both in a molecular lens and for co-magnetometry in the ACME experiment.

\section{\label{sec:setup}Experimental setup}
The measurements are carried out with cold ThO molecular beams produced by a cryogenic buffer-gas cooling source, which is a generic method for producing cold molecules~\cite{Hutzler2012,Wu2016,Singh2018,Truppe2018}. The apparatus uses neon buffer gas at $17\,$K and $40\,$sccm (standard cubic centimeters per minute) flow rate to cool pulses of ThO molecules (produced by laser ablation at $50\,$Hz repetition rate), which are then extracted into a molecular beam with $\sim\!2\,$ms pulse duration. The beam was previously characterized to have $\approx\!200\,$m/s mean forward velocity, and $\sim\!4\,$K rotational and longitudinal temperature~\cite{Hutzler2011}. 

All lasers intersect the molecular beam in nominally the transverse direction. For most of the measurements, the $Q$ state population (all states here have vibrational quantum number $v=0$, unless otherwise specified) is prepared by optically pumping on the $690\,$nm $X\rightarrow C$ transition (Fig.~\ref{fig:levelscheme}), from which about $9$\% of the population spontaneously decays into $Q$ (see \ref{sec:X2C_strength}). 
The $Q$ state population is detected $90\,$cm downstream by driving the $1196\,$nm $Q\rightarrow C$ transition, and monitoring $C\rightsquigarrow X$ fluorescence with a photomultiplier tube (PMT). The PMT has $2\times2\,$cm$^2$ active area, and fluorescence is imaged onto it by a pair of lenses with $10\,$cm focal length and $7.5\,$cm diameter. Two types of bandpass interference filters are used, for collecting $690\,$nm photons from the $C\rightsquigarrow X~(v=0)$ decay and $736\,$nm photons from the $C\rightsquigarrow X~(v=1)$ decay, respectively. The $736\,$nm off-diagonal decay is about $6$ times weaker than the $690\,$nm decay, but detecting this wavelength makes it possible to suppress background from scattered light associated with the $690\,$nm laser beams driving the $X\rightarrow C$ transition.

The Doppler width for the $Q\rightarrow C$ transition is set at $2\,$MHz full width half maximum (FWHM), with a molecular beam collimator. The mean forward velocity is monitored via time-of-flight between a measurement of the $X~(J=1)$ population just outside the buffer-gas cell and the fluorescence measurement at the end of the beam line. The former is based on absorption of $690\,$nm light tuned to the $X\rightarrow C,Q(1)$ line (where $Q(J)$ indicates the $J\rightarrow J'=J$ transition). In addition, we use this absorption signal to normalize long-term drifts in the fluorescence signal due to ablation yield variations in the beam source.

The $690\,$nm and $1196\,$nm excitation beams are derived from external cavity diode lasers. For measurements to characterize the $Q$ state, the laser frequencies are stabilized to about $1\,$MHz by a slow transfer lock to a scanning confocal cavity. The frequency tuning is calibrated based on the laser transmission through the cavity. For the STIRAP lasers, anti-reflection coated laser diodes are used to suppress fast laser phase noise, and their frequencies are stabilized to $\sim\!200\,$Hz by the Pound-Drever-Hall scheme~\cite{Drever1983} with a high finesse ultralow expansion cavity. The STIRAP laser frequencies are tuned precisely with a double-pass acousto-optic modulator system to provide tunable offsets from the cavity resonances.

\section{\label{sec:tauQ}Lifetime of the $Q$ state}

We measure the $Q$ state lifetime using a pump-probe scheme. Population in the $Q$ state is monitored with a fixed $Q\rightarrow C$ probe laser at the end of the beamline, while the $Q$ state is prepared (via $X\rightarrow C\rightsquigarrow Q$ optical pumping) at two alternating locations, $20\,$cm and $90\,$cm upstream from the probe. Fluorescence signals corresponding to these pump locations are referred to as $S_{\rm near}$ and $S_{\rm far}$, respectively. The $70\,$cm difference between the pump-probe distances corresponds to a $\delta T=3.2\,$ms difference in free flight time (with a measured mean forward velocity of $218(3)\,$m/s). The ratio between the probe signals is related to the state lifetime $\tau_Q$ by $S_{\rm far}/S_{\rm near}=\exp(-\delta T/\tau_Q)$. 

Here, optical pumping on the $X\rightarrow C, P(2)$ line (where $P(J)$ indicates the $J\rightarrow J'=J-1$ transition) is used to populate the $Q$ state. To fully saturate the pumping from all $X~(J=2)$ sublevels, at each pumping location $17\,$mW of $690\,$nm light passes through the molecule beam $7$ times, switching between orthogonal linear polarizations after each pass. Since the $P(2)$ line only has $25$\% branching ratio for decaying back to the initial state ($X,~J=2$), a short interaction time is sufficient to empty the initial state with high probability. For the $Q\rightarrow C$ probe, we use the $R(2)$ line (where $R(J)$ indicates the $J\rightarrow J'=J+1$ transition). $J=2$ is the lowest $J$ level in $Q$, and the only one populated via $C~(J'=1)$ decay. Using the $R(2)$ transition has the advantage of no dark Zeeman sublevels. This avoids potential systematic errors related to mixing of population into undetected $M$ levels via the Earth's magnetic field.

\begin{table}
\centering
\caption{\label{lifetime}Measurement of the $Q$ state lifetime. The location of the $Q\rightarrow C$ probe is fixed, while the location of the $X\rightarrow C$ pumping is alternated between $90\,$cm and $20\,$cm upstream from the probe. The values in parentheses represent the standard error of the weighted mean in the probe signal over $10$ groups of $\approx\!512$ pulses of molecules. The statistical uncertainty in the signal ratio is based on each measurement's uncertainty, inflated by the square root of the reduced $\chi^2$ values to account for the excess noise present in the data. The systematic uncertainty is discussed in the main text.}
\begin{indented}
\item[]\begin{tabular}{r c c c} \hline 
Pump-probe distance [cm]    &   &    $90$ & $20$ \\ \hline
Time of flight pump-probe [ms]    &   &    $4.1$ & $0.92$ \\ \hline
Probe fluorescence signal [a.u.]    &   &    $210(2)$ & $211(2)$ \\  \hline
reduced $\chi^2$     &    &   $1.5$ & $2.6$ \\ \hline 
signal ratio ($S_{\rm far}/S_{\rm near}$)  &    &    \multicolumn{2} {c} {$0.995\pm0.019_{\rm stat.}\pm0.021_{\rm syst.}$} \\ 
\hline \hline
$\tau_Q$ [ms]   &    &    \multicolumn{2} {c} {$>62$ ($90$\% c.\,l.)} \\
\hline
\end{tabular}
\end{indented}
\end{table}

The result is summarized in Table~\ref{lifetime}. The $Q$ state population, proportional to the probe fluorescence signal, is in fact the same between the two optical pumping locations, within the quoted statistical uncertainty. A primary contribution to the systematic error in this measurement comes from mixing between the opposite parity levels within the $\Omega$-doublet of $Q~(J=2)$ by stray electric fields between the pump and the probe regions. Depending on the details in the magnitude and direction of the stray fields over the path of the molecules, a difference in the population mixing between the two pumping locations could occur. Since the probe region has nominally zero electric field, reading out these opposite parity levels in $Q~(J=2)$ requires a $300\,$MHz frequency shift in the probe laser, corresponding to the $\Omega$-doublet spacing of the excited $C~(J=3)$ state. To place an upper bound on the effect of population mixing, we perform auxiliary measurements probing the populations of both parity levels in the $Q~(J=2)$ state. We find that the $S_{\rm far}/S_{\rm near}$ ratios are identical for both states within $2$\% uncertainty. This implies that the observed parity mixing is the same for both pump-probe distances. This is consistent with our knowledge of the $Q$ state $\Omega$-doublet (i.e., a tiny electric field is sufficient to fully mix the states; more details are provided in Section~\ref{sec:De}). After adding the statistical and systematic uncertainties in quadrature, we place a lower bound $\tau_Q>62\,$ms with $90$\% confidence. This result ensures that any $Q$ state population lives sufficiently long for our purpose, with negligible loss due to decay.

\section{\label{sec:De}$Q$ state molecule-frame electric dipole moment}

To determine the $Q$ state molecule-frame electric dipole moment $d_Q$, we prepare the population via optical pumping on the $X\rightarrow C, Q(1)$ line (note this is different from the pumping procedure in Section~\ref{sec:tauQ}.) and measure Stark shifts in the $Q\rightarrow C, P(2)$ transition. A homogeneous and variable electric field $\mathcal{E}$ is applied to the molecule detection region. The differential Stark shift is then mapped out as a function of $\mathcal{E}$ (Fig.~\ref{fig:starkmap}). The splitting of the $\Omega$-doublet in the $C~(J=1)$ state was previously measured~\cite{ESA65,Kirilov2013} to be $\Delta_C\approx51\,$MHz, and the ordering of this parity ($\mathcal{P}$) doublet was determined to have the top (bottom) level being $\mathcal{P}=-1~(\mathcal{P}=+1)$~\cite{SpaunThesis2014}. The linear polarization of the $1196\,$nm laser is nominally aligned with the $\mathcal{E}$-field and hence only excites $\pi$-transitions.
Depending on the strength of $\mathcal{E}$, the following qualitative behaviors are expected in the spectra:

 \begin{figure*}[t]
	\centering
		\includegraphics[width=0.99\textwidth]{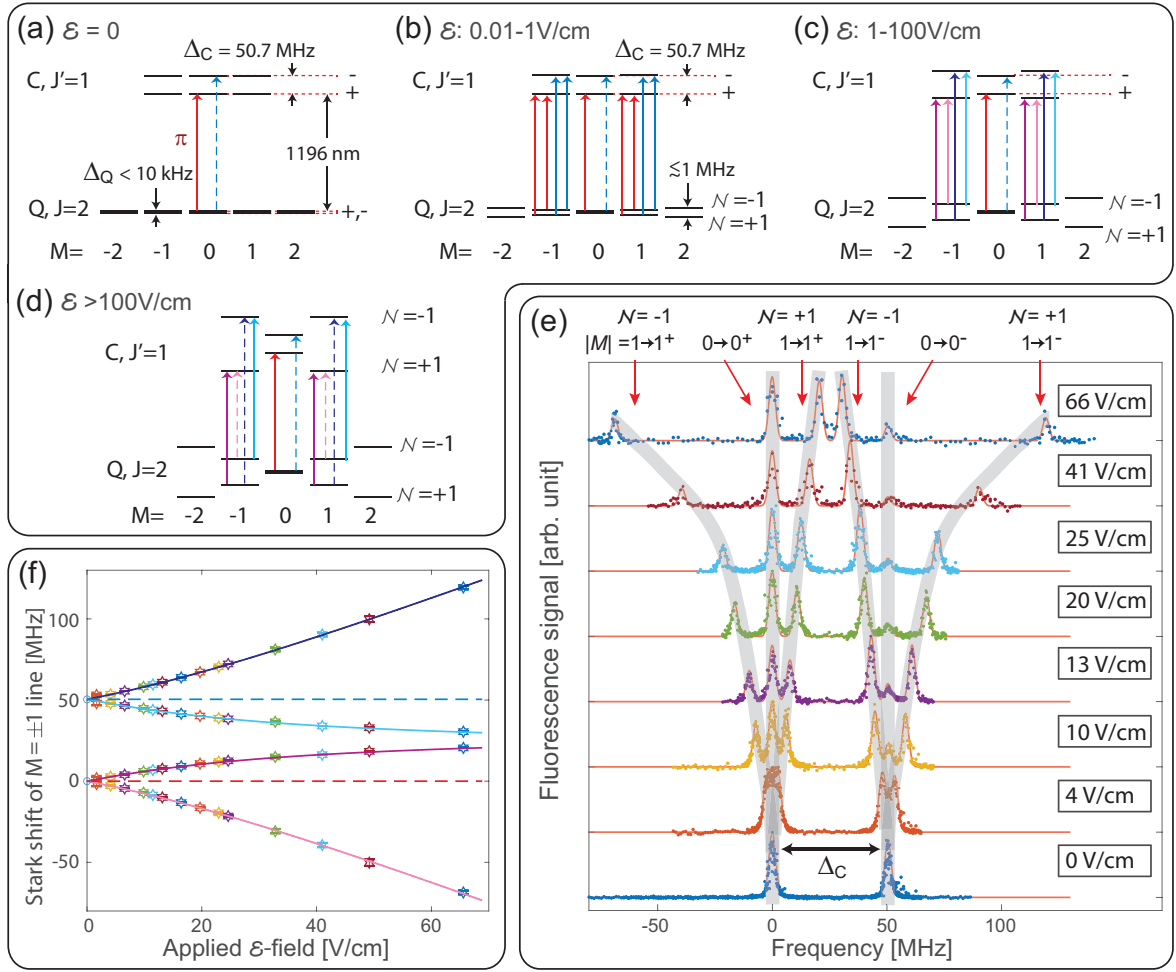}
	\caption{Differential Stark shifts in the $Q\rightarrow C,P(2)$ transition of ThO, driven with $\pi$-polarization. {\bf(a-d)} Level schemes for Stark-shifted $Q~(J=2)$ and $C~(J'=1)$, at various ranges of applied field $\mathcal{E}$. The dashed arrows indicate either forbidden or strongly suppressed transitions. The $\Omega$-doublet states of $Q~(J=2)$ have opposite parities, but these states are not resolved here. Since the $Q$ population is originally prepared only in the $\mathcal{P}=-1$ parity level, the $M=0\rightarrow M^{\prime\mathcal{P}}=0^-$ line is nominally forbidden. {\bf(e)} Spectra of $Q\rightarrow C,P(2)$ lines ($\pi$-polarization) at various $\mathcal{E}$. Solid red lines are fits to determine peak locations, using Gaussian line shapes at fixed FWHM of $3.3\,$MHz (measured at $\mathcal{E}=0$ applied). The transition to $C~(M^{\prime\mathcal{P}}=0^-)$ is still visible, but at much smaller strength. The gray bands are guides to the eye. {\bf(f)} Measured differential Stark shifts vs. $\mathcal{E}$, with theory curves fit with $d_Q$ and $d_C$ as adjustable parameters. All data are extracted from fits similar to those shown in {\bf(e)}, and error bars are all $<0.7\,$MHz. The color of the theory curves follows the same code as the transition lines in {\bf(a-d)}.}
	\label{fig:starkmap}
\end{figure*}
\begin{itemize}
\item At strictly $\mathcal{E}=0$, only a single peak should appear, associated with transition $Q~(J^\mathcal{P} = 2^-) \rightarrow C~(J^\mathcal{P} = 1^+)$ (Fig.~\ref{fig:starkmap}(a)). This is because only the $\mathcal{P}=-1$ parity level of $Q$ should be prepared via the optical pumping $X~(J^\mathcal{P}=1^-)\rightarrow C~(J^{\prime\mathcal{P}}=1^+)\rightsquigarrow Q~(J^{\prime\prime\mathcal{P}}=2^-)$ in the beamline, where (nominally) no electric field is present. 

\item At small $\mathcal{E}$ ($0.01\textendash1\,$V/cm), the $|M|=1$ sublevels in the $Q$ state get fully polarized into states either aligned ($\mathcal{N}=+1$) or anti-aligned ($\mathcal{N}=-1$)~\cite{Bickman2009,Vutha2010} with the $\mathcal{E}$-field, and experience a linear Stark shift of $\lesssim\!0.5\,$MHz. These polarized states are even and odd superpositions of the $\mathcal{P}=\pm1$ parity eigenstates. The $C$ state, on the other hand, experiences negligible parity mixing and hence a tiny quadratic Stark shift. Thus, new lines corresponding to transitions to the $C~(J^{\prime\mathcal{P}}=1^-)$ state should appear now (Fig.~\ref{fig:starkmap}(b)). For transitions to each $C~(J^{\prime\mathcal{P}})$ level with $|M'|=1$, two lines are possible: one from $Q~(\mathcal{N}=+1)$ and the other from $Q~(\mathcal{N}=-1)$. The $M=0$ level does not mix or shift, so we do not expect a new line associated with this state. Overall, since the Stark splitting of $Q$ in this regime is still too small to resolve compared to the $2\,$MHz FWHM Doppler width of the molecule beam, we expect to observe two peaks with spacing corresponding to $\Delta_C$.

\item At intermediate $\mathcal{E}$ ($1\textendash100\,$V/cm), the quadratic Stark shift of $C$ becomes visible (Fig.~\ref{fig:starkmap}(c)). The same set of lines as described in the previous scenario is expected. Here, however, the splitting between lines from the $\mathcal{N}=\pm1$ levels in $Q$ to each $C~(J^{\prime\mathcal{P}})$ state with $|M'|=1$ become fully resolved. The $Q~(M=0)\rightarrow C~(M^{\prime\mathcal{P}}=0^-)$ line remains forbidden. Thus, $5$ distinct lines should appear in the spectrum.

\item At large $\mathcal{E}$ ($\gtrsim\!100\,$V/cm), the $C$ state also becomes fully polarized into $\mathcal{N}=\pm1$ states, and transitions between opposite $\mathcal{N}$ become strongly suppressed in this regime (Fig.~\ref{fig:starkmap}(d)). Our scan does not extend to this regime, but the spectrum at $\mathcal{E}=67\,$V/cm in Fig.~\ref{fig:starkmap}(e) approaches this limit and shows reduced amplitudes for the two outermost lines.
\end{itemize}

Figure~\ref{fig:starkmap}(e) plots the Stark-shifted lines measured at various applied $\mathcal{E}$-field magnitudes. With $\mathcal{E}=0\,$V/cm (nominally), a double peak with spacing of $\Delta_C = 50.7(5)\,$MHz is observed. This agrees well with the literature value of the $\Omega$-doublet splitting of $C~(J=1)$~\cite{ESA65,Kirilov2013}. Recall that only a single peak is expected at $\mathcal{E}=0$. One likely contribution is that there exists a residual non-zero $\mathcal{E}$-field somewhere between the pump and probe regions, even with no applied voltage difference across the field plates (e.g. due to patch potentials in the surroundings). Since only $\sim 10\,$mV/cm is required to fully mix the $M=\pm1$ sublevels in the $\Omega$-doublet of $Q~(J=2)$, this seems plausible. This explanation is also consistent with behavior observed in the course of the $Q$ lifetime measurement (Section~\ref{sec:tauQ}), where full parity mixing in $Q~(J=2)$ appears to occur between the pump and probe region. The uncertainty in $\mathcal{E}$ associated with such patch potentials is, however, negligible during the rest of the Stark shift measurement, where $\mathcal{E}\gg10\,$mV/cm is applied.

Another peculiar feature in Figure~\ref{fig:starkmap}(e) is that, despite being parity forbidden, the $Q~(M=0)\rightarrow C~(M^{\prime\mathcal{P}}=0^-)$ transition is visible, although with much weaker amplitude than those lines which we expect to appear. This implies that a small amount of  $Q~(M^\mathcal{P}=0^+)$ population is produced somewhere between the pump and probe region. For molecules free-flying in this region, the quantization axis is defined dominantly by the Earth's magnetic field. By contrast, the quantization axis in the detection region is defined by the applied $\mathcal{E}$-field. In the transition between these two regions, a level crossing between the $M=0$ and $M=\pm1$ of the $\mathcal{P}=+1$ state is likely to take place, and can lead to nonadiabatic mixing of these levels. We expect this to be the main mechanism to produce population in the $Q~(M^\mathcal{P}=0^+)$ level. Nevertheless, this feature does not affect the Stark shift measurement, and provides the fit with an additional `anchor' line that does not shift with the applied field strength.

In the spectra taken at the intermediate $\mathcal{E}$ range in Fig.~\ref{fig:starkmap}(e), transitions to each level of $C~(J^{\prime\mathcal{P}})$ split into three lines, corresponding to the $Q~(M=0)$ level (which is not mixed or shifted) and the $\mathcal{N}=\pm1$ levels of $Q~(|M|=1)$. The frequency splitting is determined by the DC Stark shifts in both $Q$ and $C$. The linear Stark shift of $Q~(J=2)$, $\Delta W_Q$, is given by~\cite{Schawlow75} 
\begin{equation}
\Delta W_Q= d_Q\mathcal{E}\frac{M\Omega}{J(J+1)} = \pm \frac{1}{3}d_Q\mathcal{E},
\label{eq:WQ}
\end{equation} 
for $M=\pm1$ and $\Omega=2$, where $d_Q$ is the $Q$ state molecule-frame electric dipole moment. The quadratic Stark shift of $C~(J=1)$, $\Delta W_C$, is given by~\cite{Schawlow75}
\begin{equation}
\Delta W_C=\frac{\Delta_C}{2}\pm \sqrt{(\frac{\Delta_C}{2})^2+(d_C\mathcal{E}\frac{M\Omega}{J(J+1)})^2}
		   =\frac{\Delta_C}{2}\pm\frac{1}{2}\sqrt{\Delta_C^2+d_C^2\mathcal{E}^2},
\label{eq:WC}
\end{equation}
for $M=\pm1$ and $\Omega=1$, where $d_C$ is the $C$ state molecule-frame electric dipole moment. 

The molecule-frame dipole moments $d_Q$ and $d_C$ are simultaneously determined by fitting the $4$ measured differential Stark shifts ($\mathcal{N}=\pm1$ transitions to both $C~(J^{\prime\mathcal{P}})$ levels) with Equation \ref{eq:WQ} and \ref{eq:WC} (see Fig.~\ref{fig:starkmap}(f)). The results are $d_Q=4.07(6)\,$D and $d_C=2.60(4)\,$D. The uncertainty is dominated by systematic uncertainties, which come mainly from our estimates of the $\mathcal{E}$-field calibration due to uncertainty in the field plates spacing ($1$\%) and the frequency calibration error of the scanning cavity ($1$\%). We note that the value of $d_C$ was previously determined to be $2.53(7)\,$D in an independent measurement~\cite{HessThesis2014}. This agrees with our result within quoted uncertainties.

\section{\label{sec:Dm}Molecule-frame magnetic dipole moments}
To determine the magnetic dipole moment of the $Q$ state, we apply a homogeneous $\mathcal{B}$-field in the probe region with Helmholtz coils and measure the differential Zeeman shift of lines in the $Q\rightarrow C, P(2)$ spectrum. The spectra here contain information about both $g_Q$ and $g_C$. As discussed in the introduction, we expect $g_Q \approx 2$. In addition, since the $C$ state is largely a $^1\Pi_1$ state (Fig.~\ref{fig:levelscheme}) with $\Lambda=1$ and $\Sigma=0$, we expect $g_C\approx1$; we have independently determined $g_C=1.22(6)$ from Zeeman shifts in the $X\rightarrow C$ transition (see~\ref{sec:X2C}). 

\begin{figure}
	\centering
		\includegraphics[width=0.87\textwidth]{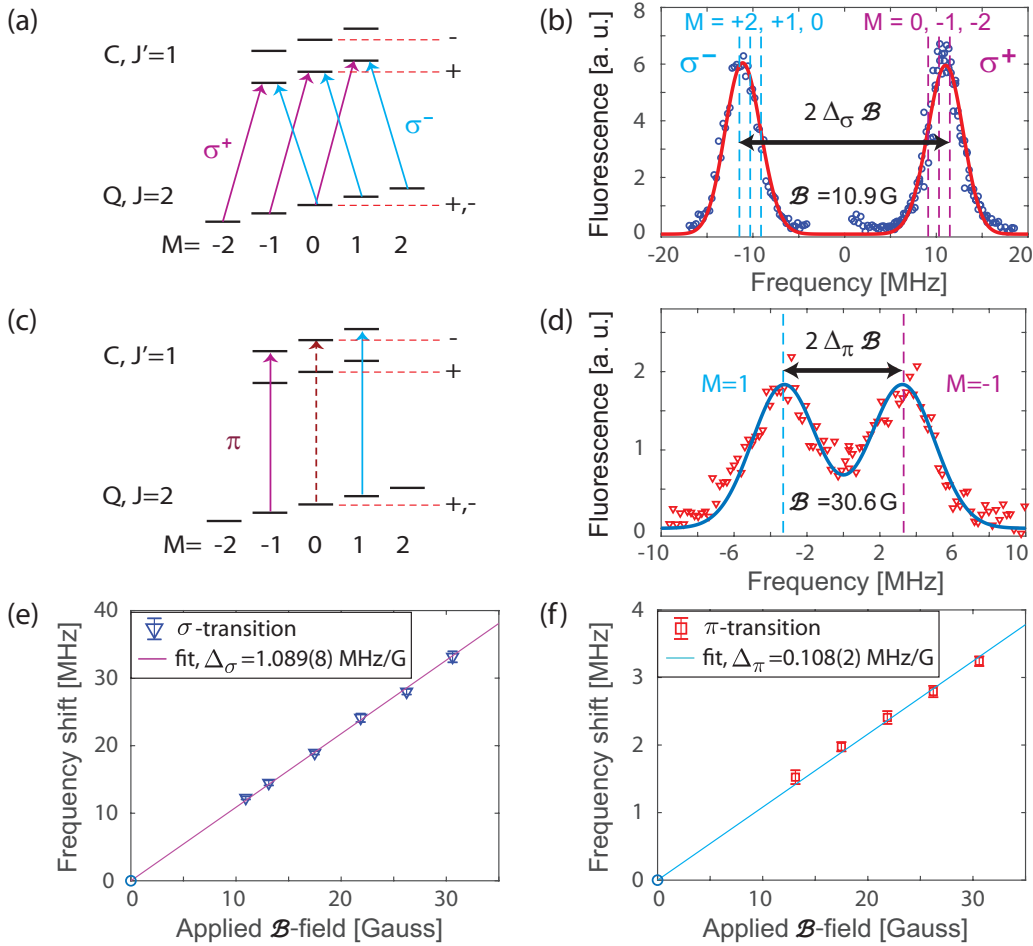}
	\caption{Differential Zeeman shifts in the $Q\rightarrow C, P(2)$ transition of ThO. {\bf(a-b)} Level diagram and a typical spectrum of Zeeman shifts with $\sigma$-polarization. For simplicity, only lines to $C~(J^\mathcal{P}=1^+)$ levels are shown. {\bf(c-d)} Level diagram and a typical spectrum of Zeeman shifts with $\pi$-polarization. For simplicity, only lines to $C~(J^\mathcal{P}=1^-)$ are shown. The line to $M^\mathcal{P}=0^-$ is nominally forbidden, as discussed in Section~\ref{sec:De}. Solid curves are fits to the line shapes. {\bf(e-f)} Linear fits to the differential Zeeman shifts in the $\sigma$ and $\pi$-transitions, respectively. The error bars in the plot and the values in parentheses indicate $1\textendash\sigma$ statistical uncertainties.}
	\label{fig:zeemanshift}
\end{figure}

The differential Zeeman shifts in $\sigma^\pm$ and $\pi$-transitions have different dependence on the $g$-factors. We fix $g_C$ with the value determined independently and obtain $g_Q$ from measurements with both polarizations. The general form for a Zeeman shift in an electronic state $\Phi$ with $\Omega\geq1$ is~\cite{Schawlow75} $\Delta W_\Phi=\mu_Bg_\Phi\mathcal{B}M\Omega/(J(J+1)),$ with $g_\Phi$ being its $g$-factor. Thus, the Zeeman shifts for $Q~(J=2,M)$ and $C~(J=1,M)$ are $\Delta W_Q=\mu_Bg_Q\mathcal{B}M/3$ and $\Delta W_C=\mu_Bg_C\mathcal{B}M/2$, respectively. Figure~\ref{fig:zeemanshift}(a) and (b) show the level scheme and a typical spectrum from $\sigma$-transitions of $Q~(J^\mathcal{P}=2^-)\rightarrow C~(J^{\prime\mathcal{P}}=1^+)$. The transitions split into two bands: $\sigma^-$ lines with negative shifts and $\sigma^+$ lines with positive shifts. The $\sigma^\pm$ band consists of three closely spaced lines from $M=(\mp2,\mp1,0)$, and the corresponding differential shifts are $\pm\mu_B\mathcal{B}\cdot(2g_Q/3-g_C/2,\,g_Q/3,\,g_C/2)$. The spacing between adjacent lines in each band is $\sim0.05\mu_B\mathcal{B}$, based on the estimated or measured value of $g_Q$ and $g_C$, and hence these adjacent lines are not resolved in the range of applied $\mathcal{B}$-field (Fig.~\ref{fig:zeemanshift}(b)). The (partially saturated) relative line strength within each band at a given probe laser power ($1\,$mW here) can be simulated accurately\footnote{The relative peak amplitudes are simulated to be $(4.1,3.6,2.5)$ for the $\sigma$-transitions from initial levels $M=(\pm2,\pm1,0)$ under our particular saturation conditions. By comparison, the relative peak amplitudes in the low power limit, given by Clebsch-Gordan coefficients squared, are $(6,3,1)$. In the opposite limit of full saturation, the peak amplitudes would be equal.} based on the model and the $Q\rightarrow C$ transition dipole moment determined in Section~\ref{sec:Dtran}. We define the splitting between the two outermost lines in the $\sigma^\pm$ band (i.e. from $M=\mp2$), which are also the strongest lines, to be $2\Delta_\sigma\mathcal{B}$, where $\Delta_\sigma=\mu_B(2g_Q/3-g_C/2)\approx0.72\mu_B$. The Zeeman splitting for $\pi$ transitions is considerably simpler (Fig.~\ref{fig:zeemanshift}(c)), given by $\pm\mu_B\mathcal{B}\cdot(g_Q/3-g_C/2,\,0)$ for $M=(\mp1,0)$. The spectrum is further simplified by driving to the $C~(J^{\prime\mathcal{P}}=1^-)$ parity level, since the $M=0$ transition is nominally forbidden as discussed in Sec.~\ref{sec:De}. This helps to resolve the small splitting in the $\pi$-transition, which we write as $2\Delta_\pi\mathcal{B}$ with $\Delta_\pi=\mu_B(g_Q/3-g_C/2)\approx0.06\mu_B$, as shown in Fig.~\ref{fig:zeemanshift}(d).

The spectrum of the $\sigma$ transitions (Fig.~\ref{fig:zeemanshift}(b)) is fit by the sum of three Gaussians for each band, with identical width based on the measured line shape at $\mathcal{B}=0$ and relative line amplitudes as discussed earlier. The spectrum of the $\pi$ transitions (Fig.~\ref{fig:zeemanshift}(d)) is fit with two symmetrically shifted Gaussians of identical width and amplitude, one for each $M=\pm1$ line. The splitting from both $\sigma$ and $\pi$ transitions are plotted in Fig.~\ref{fig:zeemanshift}(e) and (f), where it is evident that each data set fits well to a straight line. The value of $g_Q$ extracted from the two fitted linear slopes, $\Delta_\sigma$ and $\Delta_\pi$, are $2.08(2)$ and $2.06(3)$, respectively, where the values in parentheses are statistical uncertainties. We assign a final value as the weighted mean of the results, yielding $g_Q=2.07(11)$. The uncertainty is dominated by systematic contributions from our estimates of the $\mathcal{B}$-field calibration ($5$\%) due to uncertainty in the Helmholtz coil geometry, and the frequency calibration error ($1$\%) of the scanning FP cavity. 
As a consistency check, we also analyzed the spectra without constraining $g_C$, i.e. treating it as a second fit parameter. This yields $g_C=1.24(6)$, in good agreement with the value from the $X\rightarrow C$ measurement (\ref{sec:X2C}).

\section{\label{sec:Dtran}$Q\rightarrow C$ transition dipole moment}

\begin{figure}
	\centering
		\includegraphics[width=0.95\textwidth]{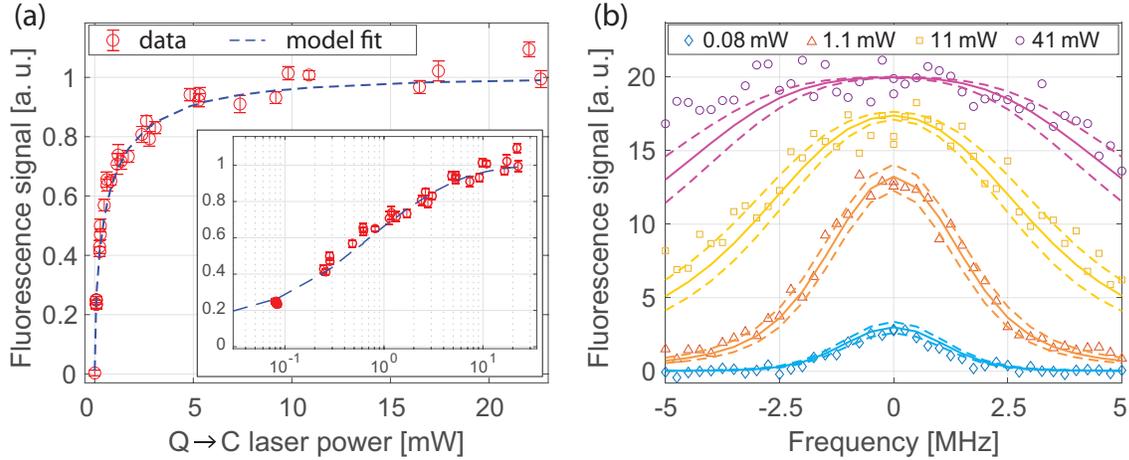}
	\caption{Data and fits to determine the transition dipole moment of $Q, |JM=2,0\rangle \rightarrow C, |J'M'=1,0\rangle$. {\bf(a)} Total fluorescence signal vs. $Q\rightarrow C$ laser power. The fluorescence signal is proportional to the integrated $C$ state population. The data are fit with the theoretical model described in the main text. The only fit parameter is the transition dipole moment $d_{Q\textendash C}$. Error bars represent the standard error in the mean over $\approx\!8$ consecutive traces, each consisting of $16$ molecule pulses.  Inset: the same data and modelled curve with the laser power plotted on a logarithmic scale. {\bf(b)} Power broadened line shapes measured at 4 different $Q\rightarrow C$ laser power levels. Solid curves are not fits, but modelled line shapes based on the transition dipole moment extracted from {\bf(a)} along with the Doppler line shape determined at the low-power limit. Dashed curves indicate the $1\textendash\sigma$ error range of the corresponding line shapes.}
	\label{fig:transitiondipole}
\end{figure}

We determine the $Q\rightarrow C$ transition dipole moment, $d_{Q\textendash C}$, by measuring the saturation intensity for this transition. To do so, we record how the integrated $C$ state population varies with the $Q, |JM=2,0\rangle \rightarrow C, |J'M'=1,0\rangle$ transition laser power (Fig.~\ref{fig:transitiondipole}(a)). An offset field of $\mathcal{E}=67\,$V/cm is applied to shift the $M=\pm1$ lines away by more than $10$ times the Doppler width. This ensures only the $|M=0\rangle\rightarrow|M'=0\rangle$ transition is driven, which simplifies the data interpretation. The data for $C$ state population versus laser power is then fit with a theoretical model which integrates the time dependent Schr\"{o}dinger equation for molecules in the process of transiting through the $Q\rightarrow C$ probe laser. This model takes into account the corresponding H\"{o}nl-London factor and Clebsch-Gordan coefficient, the longitudinal and transverse velocity distributions of the molecule beam (the latter determined from the transverse Doppler line shape), laser detuning for every transverse velocity class, and the excitation laser power and intensity profile (measured with a CCD camera). The laser beam is collimated to $3\,$mm $1/e^2$ diameter along the axis of molecule motion (horizontal, defined as $x$-axis), and expanded to $\approx18\,$mm $1/e^2$ diameter along the axis transverse to molecule beam (vertical, defined as $y$-axis). The latter is much larger than the collimated molecular beam size ($\approx5\,$mm total width) in the probe region where the molecules and laser beam intersect. Thus, molecules only fly through around the peak intensity region of the probe laser beam. The only fit parameter in the model is the transition dipole moment $d_{Q\textendash C}$. This model and the transition moment extracted from the fit are verified by independent tests of power broadened line shapes (Fig.~\ref{fig:transitiondipole}(b)). Here, line shapes measured at different laser powers are compared to the modelled line shapes using the transition dipole moment value extracted from the power saturation scan.  

We model the excitation with an open $2$-level system, with the ground state $|g\rangle=[0,1]=Q,|JM=2,0\rangle$ and excited state $|e\rangle=[1,0]=C,|J'M'=1,0\rangle$. The Hamiltonian in the rotating wave approximation is given by $H=\left[\begin{array}{cc}\delta-\Gamma&\Omega/2 \\ \Omega/2&0\end{array}\right]$, where $\delta$ is the laser detuning, $\Omega$ is the transition Rabi frequency, and $\Gamma=1/\tau_C$ is the excited state decay rate. The $C$ state lifetime $\tau_C=480(20)$\,ns was determined previously by independent methods~\cite{Kokkin2014,HessThesis2014}. The branching ratio for $|e\rangle$ decaying back to $|g\rangle$ is estimated to be only $3$\% (see Table~\ref{branching}); hence, for simplicity, the model assumes all population decays out of the system from $|e\rangle$.

A molecule starting in the $Q$ state and flying through the laser beam samples the laser intensity profile and experiences a time-dependent Rabi frequency $\Omega(t)=\mathcal{DE}(t)/\hbar$, where $\mathcal{E}(t)$ is the laser light field at time $t$, and $\mathcal{D}$ is the transition moment in the laboratory frame, related to $d_{Q\textendash C}$ via the direction cosine matrix transformation $\mathcal{D}=\mathcal{S}^{\Omega',J',M'}_{\Omega,J,M}\,d_{Q\textendash C}$. The matrix element is derived in the standard literature~\cite{Brown2003}, and in our notation yields 
\begin{eqnarray}
\fl\mathcal{S}^{\Omega',J',M'}_{\Omega,J,M}=(-1)^{\Delta M}(-1)^{M'-\Omega'}\sqrt{(2J'+1)(2J+1)}\left(\begin{array}{ccc}J&1&J' \\ -\Omega&-\Delta\Omega&\Omega'\end{array}\right)\times\nonumber\\
\left(\begin{array}{ccc}J&1&J'\\-M&-\Delta M&M'\end{array}\right).
\end{eqnarray}
Since the probe laser is continuous, $\mathcal{E}(t)=\mathcal{E}(x=t\cdot v_l,y)$ where $\mathcal{E}(x,y)$ is determined from the applied laser power and the measured intensity profile, and the $x$ and $y$ axes are defined earlier in this section. The molecule longitudinal velocity $v_l$ in the model is drawn randomly from a normal distribution centered at $220\,$m/s with FWHM of $44\,$m/s, based on measurements in~\cite{Hutzler2011}. For every value of detuning $\delta$, the probability to pump the ground state population out via the excited level after flying through the laser beam is given by $P(\delta)=1-\left\langle g\middle|\int{dt\exp(-iHt)}\middle|g\right\rangle$. Integrating $P(\delta)$ over the entire transverse Doppler distribution gives the transition probability at a given laser power in the saturation scan (dashed curve in Fig.~\ref{fig:transitiondipole}(a)). Convolving $P(\delta)$ with the transverse Doppler distribution gives the line shape curve at a given laser power (solid curves in Fig.~\ref{fig:transitiondipole}(b)).

By fitting the model to the data for fluorescence signal vs. power in Fig.~\ref{fig:transitiondipole}(a), we obtain the transition dipole moment $d_{Q\textendash C}=1.01(12)\,$D. The uncertainty is dominated by systematic errors. These include our estimated $10$\% uncertainty in the laser beam size in $y$-direction (the beam is larger than the CCD camera size, resulting in a large relative error in determination of the laser beam intensity from its power), $5$\% laser power fluctuation from the polarization-maintaining fiber that delivers the laser beam, and $3$\% power meter absolute calibration uncertainty. This value of $d_{Q\textendash C}$ is qualitatively verified by comparing the modelled power-broadened line shapes to the measured ones at different laser powers in Fig.~\ref{fig:transitiondipole}(b), using no free parameters. The good agreement between the data and model confirms the result from the power saturation fit.

\section{\label{sec:stirap}Population transfer between $X$ and $Q$ via $X\textendash C\textendash Q$ STIRAP}

\begin{figure}
	\centering
		\includegraphics[width=0.95\textwidth]{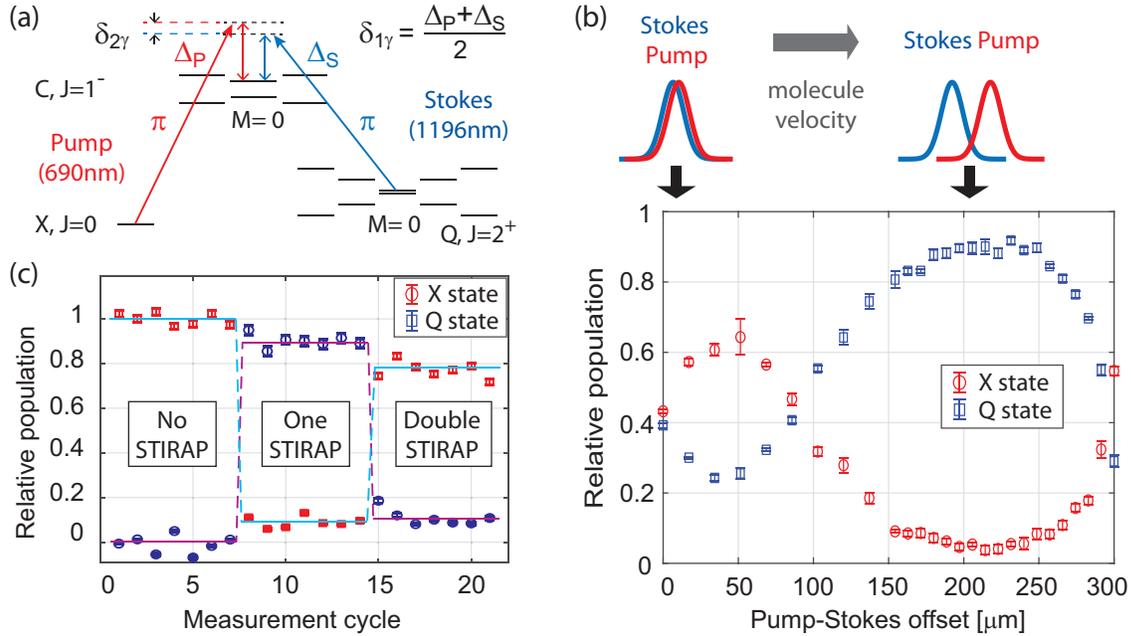}
	\caption{Demonstration of STIRAP transfer between the $X$ and $Q$ states, via the intermediate $C$ state. {\bf (a)} Level scheme for the STIRAP transfers (not to scale). The red arrow indicates the $X\rightarrow C$ Pump beam, detuned from $1$-photon resonance by $\Delta_P$. The blue arrow indicates the $Q\rightarrow C$ Stokes beam, detuned by $\Delta_S$. The $2$-photon detuning is $\delta_{2\gamma}=\Delta_P-\Delta_S$. {\bf(b)} STIRAP transfer efficiency vs. Pump-Stokes beam offset. The diagrams on top indicate the relative locations of the Pump and the Stokes beams. {\bf(c)} Population in $X,|J=0\rangle$ and $Q,|JM=2,0\rangle$ with zero, one, or two STIRAP sequences, at the optimal Pump-Stokes offset. The data in {\bf(b)} and {\bf(c)} are normalized to the population detected in $X$ with the STIRAP turned off.}
	\label{fig:stirap}
\end{figure}

The large transition moments in both $X\rightarrow C$ (see~\ref{sec:X2C_strength}) and $Q\rightarrow C$ makes it possible to efficiently transfer population between the $X$ and $Q$ state by STIRAP via the intermediate $C$ state (Fig.~\ref{fig:stirap}(a)). A similar STIRAP scheme to prepare population in the $H$ state of ThO was previously demonstrated~\cite{Panda2016}, and was used in the latest ACME measurement of the electron EDM~\cite{ACMECollaboration2018}. Using $X\textendash C\textendash Q$ STIRAP for state preparation before a molecular lens requires efficient excitation across a very large range of transverse positions and velocities (i.e. the acceptance range of the lens). The former demands high laser power to achieve sufficient intensity over the large excitation area, and the latter requires saturation of the STIRAP over a wide range of $2$-photon Doppler detunings, $\delta_{2\gamma}$. This is an extremely stringent requirement, beyond what is needed in typical applications of STIRAP~\cite{Vitanov2017}. To simulate the conditions corresponding to such a requirement in the full-scale ACME experiment, we fully open the molecule beam collimator in the laser propagation direction for these measurements. This results in a transverse velocity width of $\Delta v_{\perp}=9\,$m/s FWHM, which corresponds to a $1$-photon Doppler width (FWHM) of $13\,$MHz for the $690\,$nm Pump laser, $7.5\,$MHz for the $1196\,$nm Stokes laser, and a $2$-photon Doppler width of $5.5\,$MHz for $X\textendash C\textendash Q$ STIRAP. 

As a proof-of-principle demonstration, we drive the molecule population between $X, |J=0\rangle$ and $Q, |JM=2,0\rangle$ via $C, |JM=1,0\rangle$ (Fig.~\ref{fig:stirap}(a)). About $100\,$mW power is used in both the Pump and Stokes beams. Both lasers are focused to about $200\,\upmu$m ($1/e^2$ diameter) in the $x$ direction and collimated to about $12\,$mm ($1/e^2$ diameter) in the $y$ direction, which fully covers the $5\,$mm vertical spread of the collimated molecular beam. The Rayleigh range is $\sim3\,$cm for both laser beams, sufficient to cover the molecule beam size in the $z$ direction (along laser propagation). The $1$-photon detuning $\delta_{1\gamma}$ is deliberately set above resonance by $\approx20\,$MHz to avoid undesired resonant optical pumping due to imperfections in laser alignment, beam shape, and polarization. A constant offset electric field $\mathcal{E}=21\,$V/cm is applied continuously from the STIRAP region to the detection region, to maintain the quantization axis and to avoid transitions to unwanted $M$ sublevels. Based on the molecule-frame electric dipole moments obtained in Sec.~\ref{sec:De}, the $M=\pm1$ sublevels in $C$ are merely Stark-shifted by $3.5\,$MHz, so they stay off-resonant from undesired optical pumping by the Pump. On the other hand, the $M=\pm1$ and $\pm2$ levels in $Q$ are shifted by $14.3\,$MHz and $28.6\,$MHz, respectively, so that STIRAP into these unwanted $M$ sublevels is shifted well outside the $2$-photon Doppler width. In order to determine the relative populations in both the $X$ and $Q$ states after STIRAP, the measurement is cycled among three configurations: 1) STIRAP off, $X$-probe (via the $X,|J=0\rangle\rightarrow C,|J'M'=1,0\rangle$ line), which measures population in $X$ without STIRAP; 2) STIRAP on, $Q$-probe (via the $Q,|JM=2,0\rangle\rightarrow C, |J'M'=1,0\rangle$ line), which measures population transferred into $Q$; and 3) STIRAP on, $X$-probe again, which measures population left in $X$ after STIRAP. In all three configurations, $736\,$nm fluorescence photons from the off-diagonal decay of $C,|J'M'=1,0\rangle$ to $X\,(v=1)$ is monitored. Since the same readout state is used, this provides a direct comparison between the populations in $X$ and $Q$. In addition, to ensure the same detection efficiency for all relevant $v_\perp$-classes, both probe lasers are operated at high power to ensure strong saturation of the signal. Saturation for different velocity classes is verified both by checking that the fluorescence signal versus probe power is saturated, and with the power-broadened line shape tests, similar to those  shown in Fig.~\ref{fig:transitiondipole} and discussed in Sec.~\ref{sec:Dtran}.

A distinct feature of STIRAP is that its optimal transfer efficiency is obtained when the Stokes light field precedes the Pump field, with an offset comparable to the beam waist size~\cite{Bergmann1998}. As shown in Fig.~\ref{fig:stirap}(b), the measured state transfer efficiency indeed reaches its optimum at the expected Pump-Stokes beam offset distance, where $90$\% of the population is transferred to the $Q$ state.

To demonstrate that the STIRAP transfer works in both directions, a second set of STIRAP beams for the reversed process ($Q\textendash C\textendash X$) is also sent in $\sim1\,$cm downstream from the original $X\textendash C\textendash Q$ STIRAP. Both pairs of STIRAP beams are derived from the same laser systems, with all properties nominally identical, except the role of Pump and Stokes lasers are exchanged. As shown in Fig.~\ref{fig:stirap}(c), about $90$\% of the $Q$ population is transferred back to the $X$ state after the second STIRAP. Overall, $79(2)$\% efficiency is obtained for the ‘double'-STIRAP round-trip process.

\begin{figure}
	\centering
		\includegraphics[width=0.88\textwidth]{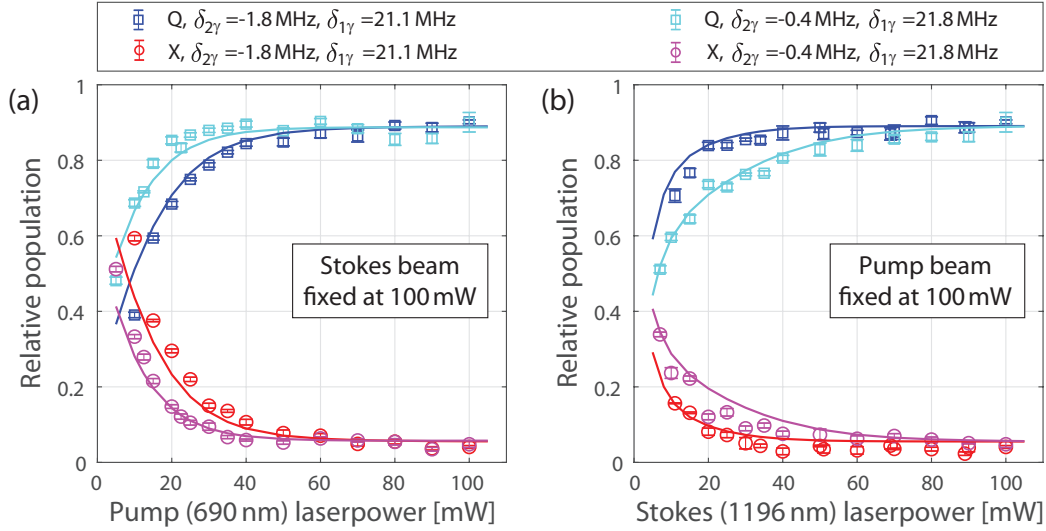}
	\caption{Population transfer versus laser power in $X\textendash C\textendash Q$ STIRAP. The $Q$ and $X$ state populations are plotted versus {\bf(a)} Pump laser power, and {\bf(b)} Stokes laser power. At each power setting, the measurement is cycled among three configurations: 1) STIRAP off, $X$-probe; 2) STIRAP on, $Q$-probe; and 3) STIRAP on, $X$-probe. Configuration 1) provides the baseline for the relative population. For each configuration, data is collected for $80$ molecule pulses. Error bars indicate the standard error in the mean over $\approx\!8$ cycles. Two values of the $2$-photon detuning $\delta_{2\gamma}$ are used. Solid curves are fits of the theoretical models described in the text. All eight curves in {\bf(a)} and {\bf(b)} share the same set of fit parameters, $\eta=0.90$ and $W_{\rm sat}=22.5\,$mW.}
	\label{fig:stirap_powerscan}
\end{figure}

To verify that the entire transverse Doppler distribution is saturated by the $X\textendash C\textendash Q$ STIRAP process, scans of the fluorescence probe signal versus STIRAP laser power are performed by varying the Pump (Stokes) beam power while keeping the Stokes (Pump) power at its maximum. Figure~\ref{fig:stirap_powerscan} shows how the STIRAP transfer efficiency saturates at $90$\% with both Pump and Stokes beams at $100\,$mW, and for two different settings of $\delta_{2\gamma}$ and $\delta_{1\gamma}$.

The data for STIRAP transfer efficiency versus laser power are fit with a theoretical model based on time-integration of Schr\"{o}dinger equation with a $3$-level system~\cite{Bergmann1998}. The Hamiltonian takes into account the realistic laser and molecular beam properties via the position-dependent Rabi frequencies and the velocity-dependent $1$- and $2$-photon detunings, as described in Section~\ref{sec:Dtran}. The transition dipole moments of the $Q\rightarrow C$ and $X\rightarrow C$  transitions in the Rabi frequencies are fixed to the values determined in Sec.~\ref{sec:Dtran} and \ref{sec:X2C_strength}, respectively. However, two corrections to this basic model are necessary in order to fully describe the data in Fig~\ref{fig:stirap_powerscan}. First, a small fraction of the population can decay out of the $3$-level system by unintended excitation to the intermediate $C, |J'M'=1,0\rangle$ state (e.g. due to suboptimal Pump/Stokes power, beam shape, or offset between the two). Of these, a ro-vibronic branching ratio of $\xi=49$\% fall back to $X, |J=0\rangle$ through  spontaneous emission. (The vibronic branching ratio is measured in \ref{sec:X2C_strength}.) This is not captured by the simple Hamiltonian model, but we resolve it by manually adding this contribution into the remaining $X$ population after STIRAP. Second, it is expected that of the population transferred into $Q$, some can get depleted by inadvertent off-resonant $Q\rightarrow C$ pumping. This can occur due to imperfections in the Stokes beam (including e.g. its back-reflection from the vacuum exit window). This process reduces the detected population in $Q$, and further increases the remaining $X$ population via spontaneous decay from $C$. To account for it, we introduce a constant rescaling factor $\eta\lesssim1$ for the detected $Q$ state population. Thus, with the populations after STIRAP simulated by the Hamiltonian model $Q_{\rm sim}$ and $X_{\rm sim}$, we expect the detected populations $Q_{\rm det}$ and $X_{\rm det}$ to be:
\begin{eqnarray}
Q_{\rm det}=\eta Q_{\rm sim}; \label{eq:Qd1}\\
X_{\rm det}=X_{\rm sim}+(1-X_{\rm sim}-Q_{\rm sim})\xi+Q_{\rm sim}(1-\eta)\xi. \label{eq:Xd1}
\end{eqnarray}
Note that here, $(1-X_{\rm sim}-Q_{\rm sim})$ is the population that gets lost from the $3$-level system after STIRAP as predicted by the basic model, and $Q_{\rm sim}(1-\eta)$ is the $Q$ state population that gets pumped out by imperfections in the Stokes beam. In the end, we fit data for $Q_{\rm det}$ and $X_{\rm det}$ with variation of the Pump power (Fig.~\ref{fig:stirap_powerscan}(a)), using $\eta$ as the only fit parameter. It makes sense to hold $\eta$ constant here since the Stokes power is kept constant during the Pump power scan.

For data of $X_{\rm det}$ and $Q_{\rm det}$ versus Stokes power (Fig~\ref{fig:stirap_powerscan}(b)), we also account for the power dependence of the $Q\rightarrow C$ pumping. We model this via the relation
\begin{equation}
Q_{\rm det}=Q_{\rm sim}(\eta+(1-\eta)e^{-W/W_{\rm sat}}),
\end{equation}
where $W$ is the Stokes laser power and $W_{\rm sat}$ is the effective saturation power. This is appropriate for a simple picture of the optical pumping as proceeding at a constant rate. The associated correction to $X_{\rm det}$ is then given by
\begin{equation}
X_{\rm det}=X_{\rm sim}+(1-X_{\rm sim}-Q_{\rm sim})\xi+Q_{\rm sim}(1-\eta)(1-e^{-W/W_{\rm sat}})\xi.
\end{equation}
Hence, when fitting the data of Fig.~\ref{fig:stirap_powerscan}(b), $W_{\rm sat}$ is the only additional unknown parameter.
At full Stokes power, the depletion of the $Q$ state population saturates at the same value of $\eta$ as in Fig.~\ref{fig:stirap_powerscan}(a), since both cases use the same maximal Stokes power. Fitting to all eight curves in Fig.~\ref{fig:stirap_powerscan} yields $\eta=0.90$, and fitting to the four curves in Fig.~\ref{fig:stirap_powerscan}(b) gives $W_{\rm sat.}=22.5\,$mW. (We note that this value corresponds to $\sim200$ times the nominal saturation intensity for the resonant $Q\rightarrow C$ transition. However, we consider this nevertheless to be a reasonable value for $W_{\rm sat.}$, since the optical pumping here only comes from imperfections in the Stokes beam.) Overall, good quantitative agreement is reached between the data and the model, for the entire power scan range and for the two different values of $\delta_{2\gamma}$.

Finally, we also recorded the dependence of STIRAP efficiency while varying the $1$-photon detunings of both the Pump ($\Delta_P$) and Stokes ($\Delta_S$) lasers when at full power. The measurement results agree well qualitatively with the modelling. Details are presented in \ref{sec:STIRAP2DScan}.

\section{\label{sec:conclusion}Conclusions and implications for ACME}
We now discuss the implications of the results for the ACME electron EDM search. In particular, we show that the $Q$ state has excellent features for building a molecular lens to enhance the statistical sensitivity, and for use as a co-magnetometer in ACME. 

First, the $Q$ state acts effectively as an additional \textquotedblleft ground\textquotedblright state of ThO molecules. We have placed a lower bound on its lifetime, $\tau_Q>62\,$ms ($90$\% confidence). This is much longer than any relevant time scale in the ACME experiment, including the time required for molecular lensing and co-magnetometry. It even offers opportunities to control the velocity of the ThO beam, using established molecule deceleration techniques~\cite{Bethlem1999,Akerman2017,Wu2017}.

Second, the $Q$ state population can be efficiently transferred from and back to the $X$ ground state, so that one can take advantage of its properties without sacrificing particle number. We demonstrate near-saturation of the $X\textendash C\textendash Q$ STIRAP transfer over the entire velocity and spatial distribution of interest in the molecule beam at $90$\% efficiency, and an $80$\% overall efficiency for the $X\textendash Q\textendash X$ round-trip population transfer.

Third, ThO molecules in the $Q$ state can be efficiently focused by either electrostatic or magnetostatic lenses. The $Q$ state molecule-frame electric dipole moment is measured to be $4.07(6)\,$Debye. It has also a near-degenerate $\Omega$-doublet of opposite parity levels. This gives rise to a strong linear Stark shift. For molecules in the $Q,|JM=2,2\rangle$ level, a trap depth of $1.8\,$Kelvin is attainable at an electric field of $\mathcal{E}=35.4\,$kV/cm. This can be realistically produced with an electrostatic hexapole lens of $19\,$mm radius, with relatively modest applied voltages of $\pm 22.5\,$kV. (At even larger $\mathcal{E}$-fields, the trap depth only increases marginally because mixing with the $J=3$ level counteracts the linear Stark shift.) The $1.8\,$K trap depth corresponds to a lens transverse capture velocity of $v_{\rm cut}\approx11\,$m/s, much larger than the acceptance velocity of $v^*_{\rm cut}\approx1.8\,$m/s obtained without a lens in the latest generation of ACME. (The latter is determined by the length of the beam line and the detection region size.) The ratio $(v_{\rm cut}/v^*_{\rm cut})^2\approx37$ sets an upper bound on the molecule flux gain from a lens. Monte-Carlo trajectory simulations for a realistic experimental geometry predict a flux gain of $\approx20$ in an optimized electric lens design. Including the $80$\% STIRAP efficiency for the $X\textendash Q\textendash X$ round-trip, an overall improvement of $16$ times in molecular signal is attainable. 

Meanwhile, we also determine the molecule-frame magnetic dipole moment to be $\mu_Q=2.07(11)\,\mu_B$. This makes it also feasible to focus ThO beams with a magnetostatic lens. A maximal field strength of $\mathcal{B}=1.5\,$T is achievable in a realistic hexapole Halbach array~\cite{HALBACH1980} of $12\,$mm inner radius, made of neodymium magnets. This yields a trap depth of $1.4\,$K for molecules in the $Q,|JM=2,2\rangle$ level. Trajectory simulations predict the flux gain from such a magnetic lens is about $50$\% lower than the electric version, due to both the lower trap depth and the smaller cross-sectional area. However, such a magnetic lens produces no difficulties with high voltage breakdown nor X-ray emission~\cite{West2017}, which might make it simpler to implement than the electric lens.

In addition, sensitive co-magnetometry using the $Q$ state could be a powerful tool to suppress $\mathcal{B}$-field related systematic errors in the ACME experiment. When prepared in a superposition of $Q,|JM=2,\pm1\rangle$ states, ThO molecules will spin-precess similar to those in the $H$ state during the EDM measurement, but at $156$ times higher sensitivity to the $\mathcal{B}$-field. The precession frequency can be read out via the $Q\rightarrow I\rightsquigarrow X$ transition\footnote{We have identified the $Q\rightarrow I, P(2)$ line at $13409.7806\,$cm$^{-1}$ with fluorescence measurement in the ThO beam line.}, analogous to how the spin precession is read out during the EDM measurement ($H\rightarrow I\rightsquigarrow X$)~\cite{ACMECollaboration2018}. Identical photon collection optics can be used for detection of both states; the only difference is that the read-out laser wavelength is $745\,$nm for $Q\rightarrow I$ versus $703\,$nm for $H\rightarrow I$.

In summary, the properties of the $Q$ state meet all the requirements for building a molecular lens of ThO, and for sensitive co-magnetometry in ACME.  A new generation of the ACME experiment is now under development, with plans to take advantage of the $Q$ state and to improve both the statistical and systematic uncertainties in the measurement of the electron EDM.

\ack This work was supported by the NSF. We thank Z. Lasner and A. D. West for helpful discussions and S. Cotreau, J. MacArthur, and M. Verma for technical support.

\clearpage
\section*{References}

\bibliography{ACMEIIbib,bib18_fixed,ACMEIIIbib,Bibliodatabase}

\clearpage

\appendix

\section{\label{sec:X2C}Independent measurement of $g_C$}
In Section~\ref{sec:Dm}, we described our measurement of the differential Zeeman shifts in the $Q\rightarrow C,P(2)$ transition. There, the frequency splitting depends on both $g_Q$ and $g_C$. Independently, we also measured the Zeeman shift in the $X\rightarrow C,Q(1)$ transition. Since the $X (^1\Sigma^+)$ state in $^{232}$Th$^{16}$O has neither an electronic contribution nor a nuclear spin, its magnetic moment should only arise from the coupling between $\mathcal{B}$ and molecular rotation~\cite{Eshbach1952}, so it is expected that $\mu_X\lesssim\mu_{\rm N}\ll\mu_C\approx1\mu_B$, where $\mu_{\rm N}$ is the nuclear magneton. Thus, the Zeeman shifts in $X\rightarrow C$ depend only on $g_C$ and the Zeeman-shifted spectra are considerably simpler to interpret than those from the $Q\rightarrow C$ transition. Comparing the value of $g_C$ extracted here with that from the $Q\rightarrow C$ transition also serves as a consistency check for the measurement in Sec.~\ref{sec:Dm}.
\begin{figure}
	\centering
		\includegraphics[width=0.8\textwidth]{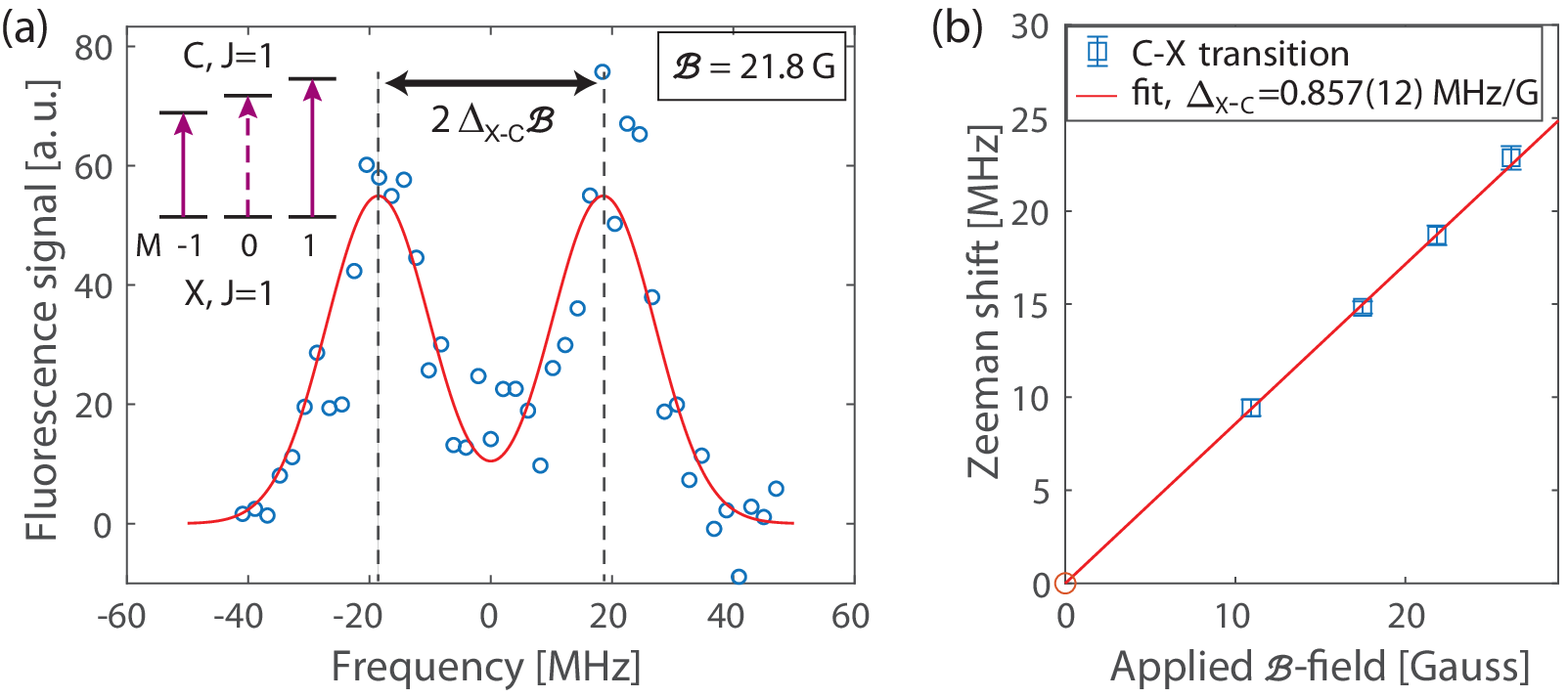}
	\caption{Zeeman splitting in the $X\rightarrow C,\,Q(1),\,\pi$-transition. {\bf(a)} A typical spectrum, taken at $\mathcal{B}=21.8\,$Gauss. The data here is noisier than that in the $Q\rightarrow C$ scan because of background scattered light from the $690\,$nm $X\rightarrow C$ probe beam. The Solid curve is a fit with two symmetrically shifted Gaussian line shapes. Inset: level diagram for the $X\rightarrow C,\,Q(1),\,\pi$-transition. $|JM=1,0\rangle\rightarrow|J'M'=1,0\rangle$ is forbidden. {\bf(b)} Zeeman shifts versus applied $\mathcal{B}$-field.  Error bars in the plots and the value in parentheses indicate $1\textendash\sigma$ statistical uncertainties.}
	\label{fig:X2C_zeeman}
\end{figure}

This measurement is performed by simply replacing the $Q\rightarrow C$ probe laser with a $X\rightarrow C$ laser. Figure \ref{fig:X2C_zeeman} presents a typical spectrum and the frequency shifts vs. applied $\mathcal{B}$. Here, the shift of a $\pi$-transition line between states with quantum number $M$ is given by $\mu_Bg_CM\mathcal{B}/2$. Thus, the splitting between lines in Fig.~\ref{fig:X2C_zeeman}(a) is $2\Delta_{X\textendash C}\mathcal{B}$, with $\Delta_{X\textendash C}=\mu_Bg_C/2$. In Fig.~\ref{fig:X2C_zeeman}(b), the value of $\Delta_{X\textendash C}$ is determined from the linear fit to data from Zeeman shifts versus $\mathcal{B}$, yielding the molecule-frame magnetic moment in the $C$ state, $g_C=1.22(6)$. The uncertainty here is dominated by the systematic uncertainty from calibration of the applied $\mathcal{B}$ field ($5$\%), due to imperfections in the Helmholtz coil geometry. This result agrees well with the value obtained in Sec.~\ref{sec:Dm} using the $Q\rightarrow C$ transition alone.

\section{\label{sec:X2C_strength}Transition dipole moment of $X\rightarrow C$ and branching ratio for decays from $C$}

\begin{figure}
	\centering
		\includegraphics[width=0.68\textwidth]{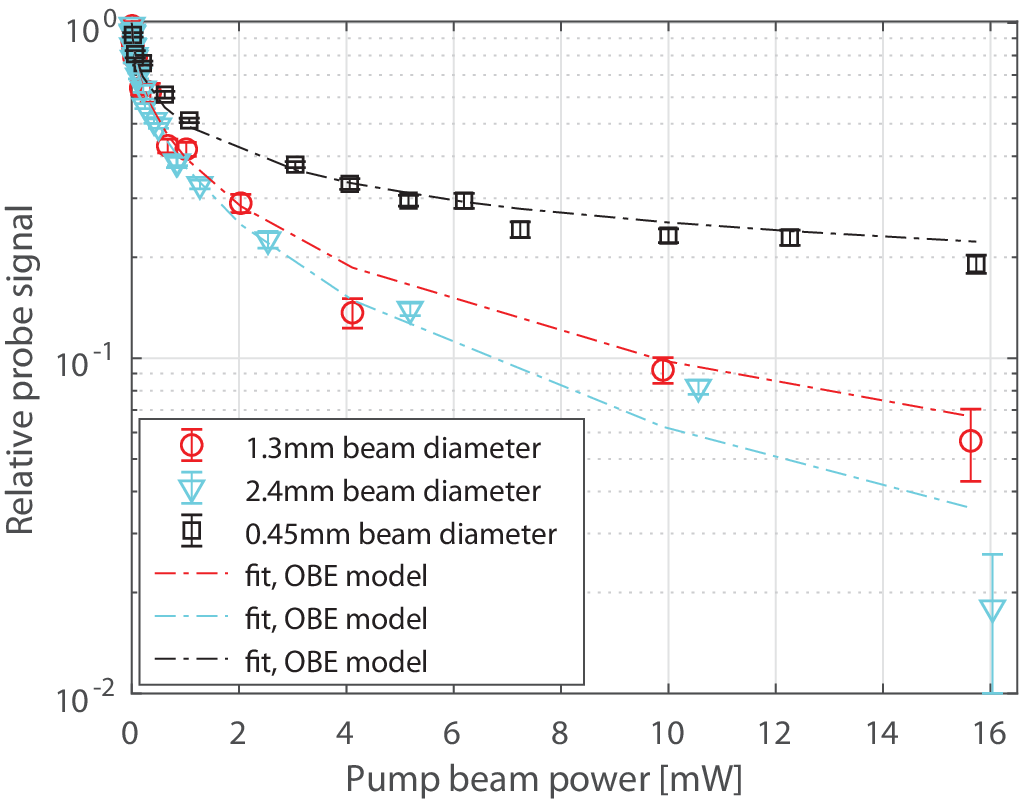}
	\caption{Pump-probe measurement on $X\rightarrow C,\,R(0), \pi$-transition. Pump power is varied, depleting population from $X,|JM=0,0\rangle$ state via optical pumping through the $C,|J'M'=1,0\rangle$ state. Probe power is fixed, and the fluorescence signal is proportional to population remaining in the ground state after pumping. Three different pump beam diameters (along the direction of molecule beam longitudinal velocity) are used. The error bars indicate the standard error in the mean over typically $8$ traces and each trace consists of $16$ molecule pulses. Dashed lines are fits from optical Bloch equation models, with the molecule-frame transition dipole moment $d_{X,0\textendash C,0}$ being the only fit parameter. The best fit values for the three measurements are given in the text.}
	\label{fig:X2C_scan}
\end{figure}

In this section, we summarize a measurement of the $X\rightarrow C$ transition dipole moment $d_{X\textendash C}$. This is accomplished by both pumping and probing on the $X,|J=0\rangle\rightarrow C,|J'M'=1,0\rangle$ line. At a fixed probe power, the dependence of the probe fluorescence signal on the pump power is measured (Fig.~\ref{fig:X2C_scan}). The pump laser depletes the population from the $X,|J=0\rangle$ state, while the probe signal measures the population remaining there. By fitting the dependence of the remaining population versus pump power to a theoretical model of the optical pumping process, the $X\rightarrow C$ transition dipole $d_{X\textendash C}$ can be determined. 

Both the measurement and the analysis here are very similar to those discussed in Sec.~\ref{sec:Dtran}, from which the $Q\rightarrow C$ transition strength was derived. The main difference in the analysis is that instead of solving the time-dependent Schr\"{o}dinger equation, here we use the optical Bloch equations to fully describe the dynamics of molecules flying through the $X\rightarrow C$ pumping beam. This is essential to accurately account for the $C, |J'M'=1,0\rangle\rightsquigarrow X, |JM=0,0\rangle$ spontaneous decay, which has a combined rotational and vibronic branching ratio $\xi$ (known {\it a priori} to be $\sim50$\%) to land in the initial state. $\xi$ can be expressed in terms of the molecule-frame transition dipole moment $d_{X,0\textendash C,0}$~\cite{Kokkin2014}
\begin{equation}
\xi=\frac{2}{3}\xi_{C,0\textendash X,0}=\frac{2}{3}\times3.137\times10^{-13}\left[\frac{{\rm m}^3}{{\rm D}^2{\rm s}}\right]\frac{d^2_{X,0\textendash C,0}}{\lambda^3_{X,0\textendash C,0}}\tau_C,
\label{eq:xiXC}
\end{equation}
where the factor of $2/3$ is the rotational branching ratio, $\xi_{C,0\textendash X,0}$ is the vibronic branching ratio for the $C~(v'=0)\rightsquigarrow X~(v=0)$ decay, $\lambda_{X,0\textendash C,0}=690\,$nm is the wavelength, and the $C~(v'=0)$ state lifetime $\tau_C=480(20)\,$ns is obtained from the weighted average of two independent measurements~\cite{Kokkin2014,HessThesis2014}. Thus, $d_{X,0\textendash C,0}$ is the only unknown parameter in our optical Bloch equation model, and is determined from the fit to data of probe signal versus pump power (Fig.~\ref{fig:X2C_scan}). The value of $\xi_{C,0\textendash X,0}$ can then be deduced from Equation~\ref{eq:xiXC}.

To estimate possible systematic errors in our obtained value of $d_{X,0\textendash C,0}$, we perform the measurement at three different waist sizes of the pump laser beam. The beam diameter ($1/e^2$) in the $x$ direction is varied from $450\,\mu$m to $2.4\,$mm, while in the $y$ direction it is kept at $13\,$mm  (much larger than the $5\,$mm collimated molecule beam size). In this way, both the excitation laser intensity and the interaction time between molecule and light are drastically changed. Fits in Fig.~\ref{fig:X2C_scan} yield $d_{C,0\textendash X,0}$ values of $1.25^{+0.02}_{-0.05}, 1.28^{+0.05}_{-0.03}$, and $1.32^{+0.04}_{-0.06}\,$Debye, corresponding to the measurements with $450\,\mu$m, $1.3\,$mm, and $2.4\,$mm beam diameters. The uncertainties quoted here are $1\textendash\sigma$ statistical errors. A weighted average gives $\bar{d}_{C,0\textendash X,0}=1.27(5)\,$D, where the uncertainty consists of a statistical contribution ($1.7$\%) from the standard error of the weighted mean, and systematic contributions including an uncertainty ($2$\%) from the $\tau_C$ measurement and an estimate based on the standard deviation ($3$\%) of $d_{C,0\textendash X,0}$ under the three different measurement conditions. The corresponding branching ratio is $\xi_{C,0\textendash X,0}=74(6)$\%.

\begin{table}
\centering
\caption{\label{branching}Branching ratios of decays from the $C~(v'=0)$ state of ThO. Branching ratios to intermediate states other than $X$, $Q$, and $H$, i.e., $A$, $B$, $H'$, $K'$, and $N'$, are estimated based on the amplitudes of their decomposition in the Hund's case (a) basis~\cite{Kaledin2019} and the wavelengths for decay to them from $C~(v'=0)$. The uncertainty of these estimated values can be as large as a factor of $2$.}
\begin{indented}
\item[]\begin{tabular}{l c c} \hline \hline \hline
{\bf Vibronic state} & {\bf Branching ratio} & {\bf Source} \\ \hline \hline \hline
$\xi_{C,0\textendash X,0}$ & $74(6)$\% & $d_{C,0\textendash X,0}$ this work, $\tau_{C}$ in~\cite{Kokkin2014,HessThesis2014}\\ \hline
$\xi_{C,0\textendash X,1}$ & $13(1)$\%  &Franck-Condon factors in~\cite{WS72}\\ \hline
$\xi_{C,0\textendash X,2}$ & $1.0(1)$\%  &Franck-Condon factors in~\cite{WS72}\\ \hline \hline \hline
$\xi_{C,0\textendash Q,0}$ & $9(2)$\%  & $d_{C,0\textendash Q,0}$ this work, $\tau_{C}$ in~\cite{Kokkin2014,HessThesis2014}\\  \hline
$\xi_{C,0\textendash Q,1}$ & $1.0(2)$\%  & Franck-Condon factors in~\cite{Kokkin2014}\\  \hline\hline \hline
$\xi_{C,0\textendash H,0}$ & $0.04(2)$\%  & $d_{C,0\textendash H,0}$ in~\cite{SpaunThesis2014}, $\tau_{C}$ in~\cite{Kokkin2014,HessThesis2014}\\  \hline
$\xi_{C,0\textendash A,0}$ & $0.2(2)$\%  & estimate in~\cite{HessThesis2014}\\  \hline
$\xi_{C,0\textendash B,0}$ & $0.8(8)$\%  & estimate in~\cite{HessThesis2014}\\  \hline
$\xi_{C,0\textendash H',0}$ & $0.4(4)$\%  & estimate based on~\cite{Kaledin2019}\\  \hline
$\xi_{C,0\textendash K',0}$ & $0.2(2)$\%  & estimate based on~\cite{Kaledin2019}\\  \hline
$\xi_{C,0\textendash N',0}$ & $0.1(1)$\%  &estimate based on~\cite{Kaledin2019}\\  \hline
\hline \hline
{\bf Sum} & ${\bf 100(6)\%}$ &uncertainties added in quadrature\\
\hline \hline\hline
\end{tabular}
\end{indented}
\end{table}

Due to the large transition dipole moment of $X\rightarrow C$ and the availability of high quality laser diodes and tapered amplifiers at this wavelength, the $C$ state of ThO has been used in a variety of ways for ACME, including for rotational state cooling, STIRAP population transfer, and fluorescence detection. Thus, it is of interest to understand the branching ratios of various decays from $C$. From the transition dipole moment $d_{Q,0\textendash C,0}=1.01(12)\,$D determined in Sec.~\ref{sec:Dtran}, together with the values of $\tau_C$ and $\lambda_{Q,0\textendash C,0}$, we obtain the $C~(v'=0)\rightsquigarrow Q~(v=0)$ vibronic branching ratio, $\xi_{Q,0\textendash C,0}=9(2)$\%. The $H~(v=0)\rightarrow C~(v'=0)$ transition dipole moment was measured in~\cite{SpaunThesis2014}: $d_{H,0\textendash C,0}=0.06(3)\,$D, with a wavelength of $\lambda_{H,0\textendash C,0}=1090\,$nm. Thus, we can calculate $\xi_{H,0\textendash C,0}=0.04(2)$\%. In addition, the Franck-Condon factors of $C~(v'=0)\rightsquigarrow X~(v=0,1,2)$ and $\rightsquigarrow Q~(v=0,1)$ are known~\cite{WS72,Kokkin2014}. The molecular constants for all relevant electronic states of ThO, summarized in~\cite{Edvinsson1985}, can be used to calculate Franck-Condon factors for decays from the $C~(v'=0)$ state.  From the ratio of the Franck-Condon factors, we can calculate the branching ratios for the off-diagonal decays of $C\rightsquigarrow X$ and $C\rightsquigarrow Q$. Table~\ref{branching} summarizes the branching ratios for all known possible decays from the $C~(v'=0)$ state, and shows they account for $100$\% to within our uncertainties.

\section{\label{sec:STIRAP2DScan}$X\textendash C\textendash Q$ STIRAP efficiency vs. $\delta_{2\gamma}$ and $\delta_{1\gamma}$ detuning}
\begin{figure}
	\centering
		\includegraphics[width=0.99\textwidth]{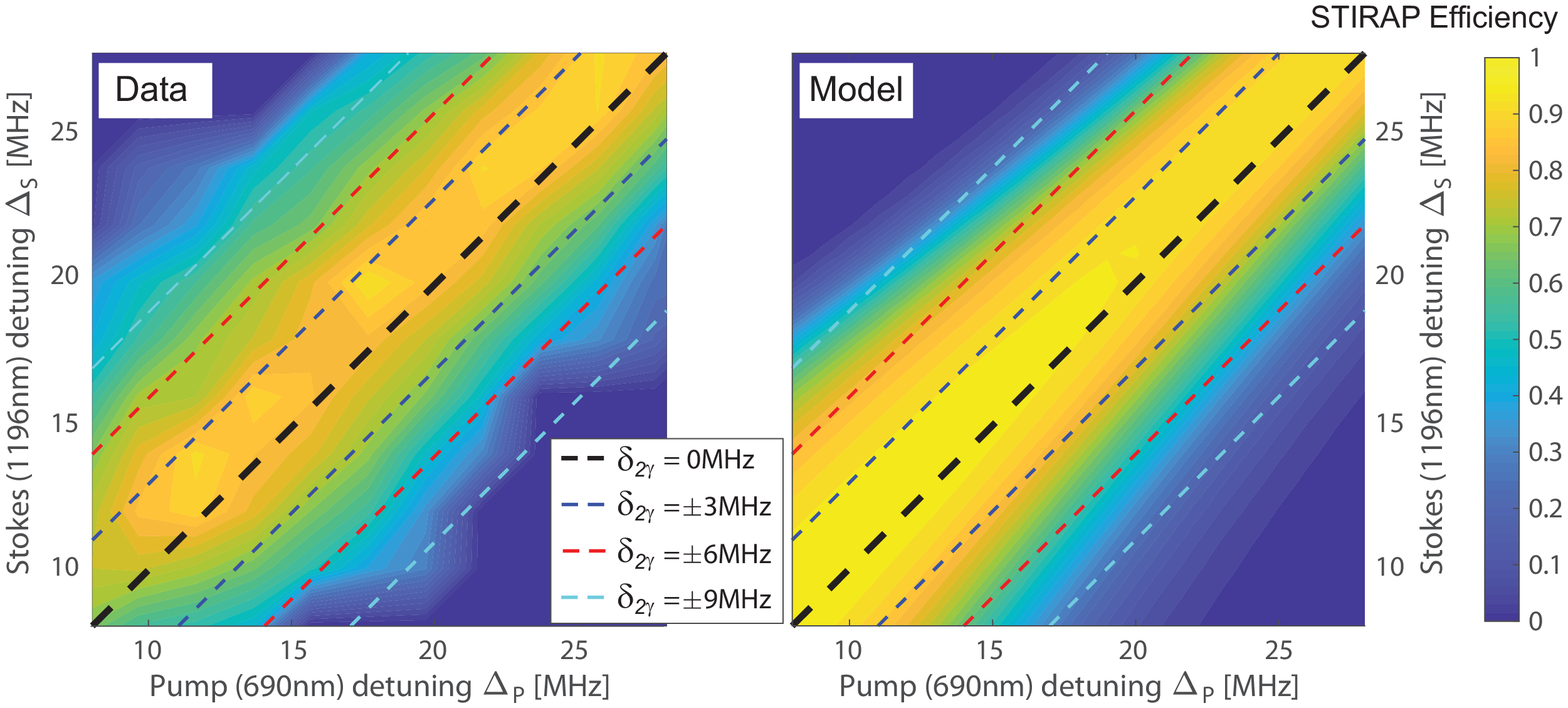}
	\caption{$X\textendash C\textendash Q$ STIRAP efficiency vs. detuning of Pump and Stokes lasers. The transition level scheme is given in Fig.~\ref{fig:stirap}(a). $2$-photon detuning is defined as $\delta_{2\gamma}=\Delta_P-\Delta_S$. The plot on the left is experimental data. The measurement conditions are given in the text. The plot on the right is from simulations based on the Hamiltonian of the $3$-level system, as outlined in Sec.~\ref{sec:stirap}. }
	\label{fig:stirap2D_scan}
\end{figure}

In the course of optimizing the $X\textendash C\textendash Q$ STIRAP efficiency, we varied the detuning of both the Pump ($\Delta_P$) and Stokes ($\Delta_S$) lasers. The level diagram for the STIRAP process and the definitions of $\delta_{2\gamma}$ and $\delta_{1\gamma}$ are given in Fig.~\ref{fig:stirap}(a). In Fig.~\ref{fig:stirap2D_scan}, the measured  STIRAP efficiency vs. $\delta_{2\gamma}$ is compared with results from modelling. The measurement is taken with $120\,$mW Pump power and $290\,$mW Stokes power. Both laser beam waists are softly focused to $420\,\mu$m $1/e^2$ diameter. The Pump-Stokes beam offset is about $420\,$um. The model is the same as outlined in Sec.~\ref{sec:stirap}, where the Hamiltonian of the $3$-level system is integrated as molecules transit through the STIRAP laser beams.

The data and modelling in Fig.~\ref{fig:stirap2D_scan} agree well qualitatively. An interesting feature, clearly present in both the data and the model, is that the STIRAP efficiency is asymmetric with respect to the $2$-photon resonance. In both plots, the optimum efficiency is reached at $\delta_{2\gamma}$ between $0$ and $-3\,$MHz. This asymmetry was investigated in~\cite{Boradjiev2010}. It can be understood intuitively as follows. Since the STIRAP is deliberately tuned off the $1$-photon resonance, the AC Stark shifts are in the same direction for both the Pump and Stokes beams. In this case, a difference in the Rabi frequencies between $X\rightarrow C$ and $Q\rightarrow C$ causes an imbalance in the AC Stark shifts induced on the $X$ and $Q$ states. This leads to a shift in the $2$-photon detuning.

\clearpage

\end{document}